\documentclass[
reprint,
aps
]{revtex4-2}

\usepackage{amsmath}
\usepackage{amssymb}
\usepackage{graphicx}% Include figure files
\usepackage{dcolumn}% Align table columns on decimal point
\usepackage{bm}% bold math
\usepackage{hyperref}% add hypertext capabilities
\usepackage{comment}
\usepackage{xcolor}
\usepackage{framed}

\usepackage{CJKutf8}

\usepackage{tikz}
\usetikzlibrary{decorations.pathmorphing}

\allowdisplaybreaks  

\newcommand{\bk}[0]{\mathbf{k}}
\newcommand{\bp}[0]{\mathbf{p}}
\newcommand{\bx}[0]{\mathbf{x}}
\newcommand{\bfxi}[0]{\boldsymbol{\xi}}
\newcommand{\bfeta}{\boldsymbol{\eta}}
\newcommand{\bka}[1]{\left\langle #1 \right\rangle}
\newcommand{\br}{\mathbf{r}}
\newcommand{\calD}{\mathcal{D}} %functional differential

\newcommand{\dif}{\mathrm{d}}
\newcommand{\ext}{\mathrm{ext}}
\newcommand{\eff}{\mathrm{eff}}

\begin{document}
\title{Effective diffusion of a tracer in active bath: a path-integral approach} 
\author{Feng, Mengkai}
\email{fengmk@ustc.edu.cn}
\author{Hou, Zhonghuai}
\email{hzhlj@ustc.edu.cn}
\affiliation{%
 Hefei National Research Center for Physical Sciences at the Microscale \\
 \& Department of Chemical Physics, University of Science and Technology of China, Hefei, Anhui 230026, China
}%
\date{\today}

\begin{abstract}
We investigate the effective diffusion of a tracer immersed in an active particle bath consisting of self-propelled particles.
Utilising the Dean's method developed for the equilibrium bath and extending it to the nonequilibrium situation, we derive a generalized Langevin equation (GLE) for the tracer particle.
The complex interactions between the tracer and bath particles are shown as a memory kernel term and two colored noise terms.
To obtain the effective diffusivity of the tracer, we use path integral technique to calculate all necessary correlation functions. 
Calculations show the effective diffusion %has a linear relationship with the amplitude of the active force when the activity is small, 
decreases with the persistent time of active force, and has rich behavior with number density of bath particles, depending on different activity. 
All theoretical results regarding the dependence of such diffusivity on bath parameters have been confirmed by direct computer simulation.

%GPT rewrite
%We explore the intricacies of effective diffusion exhibited by a tracer immersed within an active particle bath. Employing the Dean's method initially devised for equilibrium baths and extending its application to non-equilibrium scenarios, we derive a generalized Langevin equation (GLE) specifically tailored for the tracer particle. This GLE accounts for the complex interplay between the tracer and bath particles, manifesting as a memory kernel term and two colored noise terms. To ascertain the effective diffusivity of the tracer, we leverage path integral techniques for calculating the necessary correlation functions. Our numerical computations illuminate a linear relationship between effective diffusion and the amplitude of the active force in scenarios where activity remains modest. Additionally, we observe a reduction in effective diffusion with increasing persistent time of the active force and note the diverse behavior tied to the number density of bath particles, contingent upon varying activity levels. Theoretical predictions regarding the diffusivity's dependence on bath parameters are further validated through direct computer simulations, offering qualitative alignment with our theoretical framework.
\end{abstract}

\maketitle

\section{Introduction}
Active matter systems, consisting of self-propelled units that are
able to convert stored or surrounding energy into their persistent motion, provide a fresh opportunity for applications of nonequilibrium statistical mechanics \cite{00PhyA_Veseck, 13RevModPhys_HydroAct, 15RPP_Elgeti_SingleAndCollective, 16RMP_BechingerActive}.
In particular, active colloidal suspension can serve as an active bath that can significantly influence the motion and dynamics of passive objects submerged within them \cite{00PRL_Wu_bacteria_bath, 04PRL_bacteria, 16PRL_DiffBacSus, 20_PRL_YangMC, 20SM_YMC_noise, 22SM_YMC_noise}.
Understanding the behavior of tracer particles in the active bath is a fundamental pursuit in statistical physics and plays a crucial role in various biological\cite{17PRE_nonGauss, 18Bio_ArigaNonequilibrium, 20BioArigaExperimental,23PRL_YMC_HI-BS}, chemical\cite{18chemotaxis, 23PRL_chemotactic}, and physical phenomena\cite{16JSM_TracerMotionActiveDumbb, 16NatPhys_HeatEngineBacteriaBath, 17SM_HeatEngine, 17PRE_BradyTracerDiffActBath, 19JSM_StochEngine, 20PRE_BrownHE, 20_PRL_YangMC, 20SM_YMC_noise, 22SM_YMC_noise, 22PRR_NonFluRes, 22PRX_secondLaw, 23RPPFeng, 23NJP_Das, 23PRL_infoEng,23PRL_WorkFluAOUP}. 
Notably, the tracer particle immersed in the active bath exhibits a distinct diffusion profile compared to its equilibrium counterpart\cite{04PF_Kim_EnhancedDiff, 14PRL_Maggi_GenaralEnergyEquiInAB, 14JSP_Maes_2ndFDT, 15ScitiRep_Maggi_Multidimen_Stationary, 17SciRep_Maggi_MemLessResponseAndFDT, 20SM_Demery_BacSus, 22PRR_NonFluRes, 22JSM_Tracer1dBath}. 
Furthermore, studying the diffusion behavior of such a tracer is essential for unraveling the intricate dynamics that govern systems away from thermal equilibrium, providing a valuable tool to investigate the collective behavior \cite{12PhysRep_Vicsek_CollectiveMotion, 12Nature_Suminol_HardFilament, 12PRL_collectmotion, 13RMP_Marchetti_HydroSoftActi, 20PNAS_collective, 20_PRL_YangMC, 22NJP_PassToAct, 23ComPhys_jiege} and transport properties in such medias \cite{09RMP_Hanggi, 12RPP_Cates_Diffusive_transport, 14SM_transportAP, 17PNAS_transport, 22PRL_anomalous}. 
Because of the importance and wide range of applications, understanding the dynamics of the tracer in such an active bath is desirable.

As one already knows, the classical work by Einstein laid the foundation for understanding Brownian motion, providing a framework for diffusive behavior in passive media. Subsequent advancements, such as the Langevin equation, have enriched our understanding of stochastic processes, diffusive phenomena and fluctuation-dissipation theorem in thermal equilibrium. 
Nevertheless, the dynamics of particles in active baths introduce additional complexity\cite{14PRE_curvatureTracer, 15PRE_Maes_FricNoiseProbe, 17SciRep_Maggi_MemLessResponseAndFDT, 20PRL_MaesActBath, 20SM_KR_noise, 21sm_crowdedAP, 22PRL_anomalous, 22JPA_Solon_ERActBath,22NJP_PassToAct}. 
Several works have investigated this question, often employing analytical and numerical techniques to model and characterize the motion of tracer particles within nonequilibrium media including the active bath.
For instance, Maes \textit{et al.} established a generalized fluctuation-response relation for thermal systems driven out of equilibrium \cite{09PRL_MaesNonRes, 11JSM_Maes_FluRes, 13PRE_MaesFRR, 15JPCA_Maes_ResFrenesy, 20PhysRep_Maes_Frenesy, 20_FrontPhys_Maes_respon}, utilized this method to investigate the fluctuation-dissipation relation for nonequilibrium bath \cite{14JSP_Maes_2ndFDT}, and further studied the dynamics of a tracer immersed in such bath, gave the friction and noise properties \cite{15PRE_Maes_FricNoiseProbe}, the Langevin description \cite{16JPCM_Maes_Langevin}, and correlation functions of the tracer variables to study the fluctuation properties\cite{20PRL_MaesActBath}.
Speck and Seifert \textit{et al.} formulated a fluctuation-dissipation theorem (FDT) within a nonequilibrium steady state of a sheared colloidal suspension system \cite{10_EPL_Seifert_Speck_FDT, 10PRE_Bechinger_GFDR}, and subsequently investigated the mobility and diffusivity of a tagged particle within this system, determined the velocity autocorrelation functions and response functions with small shear force, found that a phenomenological effective temperature recovers the Einstein relation in nonequilibrium \cite{11EPL_SeifertSpeck_MobDiff}.
Lauga \textit{et al.} proposed a stochastic fluid dynamic model to describe analytically and computationally the dynamics of microscopic particles driven by the motion of surface attached bacteria, analytically calculated expressions for the effective diffusion coefficient through a run-and-tumble model, found that the short-time mean squared displacement is proportional to the square of the swimming speed while the long-time one only depends on the size of the particle \cite{19SM_Lopez_StoModBactDrivSwimmer}.
Brady \textit{et al.} studied the diffusion of a tracer in a dilute dispersion of active Brownian particles (ABPs), by employing the Smoluchowski equation and averaging over bath particles and orientation variables, obtained tracer’s single-particle probability distribution function, found that the active contribution to the diffusivity scales as $U_0$ (characteristic swim speed of ABP) for strong swimming and $U_0^2$ for weak swimming \cite{17PRE_BradyTracerDiffActBath}. Furthermore, they derived a general relationship between diffusivity and mobility in generic colloidal suspensions, provided a method to quantify deviations from the FDT and express them in terms of an effective SES relation \cite{19_JCP_Brady_FDT}.
More recently, Omar \textit{et al.} studied the long-time dynamics of a tracer immersed in an one-dimensional active bath, derived a time-dependent friction and noise correlation with power law long tails that depend on the symmetry of tracers, and found that shape asymmetry of the tracer induces ratchet effects and leads to super-diffusion and friction that grows with time \cite{22PRL_anomalous}. 

Numerous theories based on various starting points have demonstrated the importance and attraction of studying tracer behavior in nonequilibrium baths.
In this study we propose an alternative theoretical method based on path-integral method to investigate the behavior of tracer diffusion in an active particle bath, and subsequent simulation results successfully validate our theory.
The starting point of the theory is the generalized Langevin equation (GLE) for the tracer, by utilizing a generalized version of Dean's equation to describe the active bath. 
The GLE contains a memory kernel function and complex effective noise terms, reflecting the complex interactions between the tracer and bath particles.
We then employ the path integral method \cite{11PRE_Demery_TracerInFluids, 14NJP_GLE_Demery} to calculate the diffusion coefficient.
Numerical calculations show that the effective diffusion has a non-trivial dependence on bath parameters such as the number density and the persistent time of the active bath particle.
Finally, we perform extensive computer simulations, which show very good agreement with our theoretical predictions.%, which sufficiently enhancing the robustness of our findings.

This work is organized as following: In Sec.II, we introduce the model system and derive the GLE of the tracer. Then, we utilize the path integral method to obtain the effective diffusion of the tracer formally. In Sec.III, we show the numerical solution of such diffusion and compare it with simulation results. And the paper ends with conclusion in Sec.IV.

\section{Model and theory}
\subsection{Active bath model}
Consider a system consisting of a tracer particle and $N$ active Ornstein-Uhlenbeck(OU) particles. The coordinates of the tracer $\mathbf{x}$ and the active bath particles $\mathbf{r}_i$ are governed by the overdamped Langevin equations,
\begin{subequations}
\begin{align}
    \dot{\mathbf{x}} =& -\mu_t \nabla_x \left[ \sum_{i=1}^N U(|\mathbf{x}-\mathbf{r}_i|) + U_{\ext}(\mathbf{x}) \right] + \sqrt{2\mu_t T} \boldsymbol{\xi}_t,  \\
    \dot{\mathbf{r}}_i =& -\mu_b \left\{ \mathbf{f}_i + \nabla_i \left[ \sum_{j\neq i} V(|\mathbf{r}_i - \mathbf{r}_j|) + U(|\mathbf{r}_i-\mathbf{x}|) \right] \right\} \nonumber \\
    & + \sqrt{2\mu_b T} \boldsymbol{\xi}_i, \label{eq:LEs} \\
    \tau_b \dot{\mathbf{f}}_i =& -\mathbf{f}_i + \sqrt{2D_b/\mu_b^2} \boldsymbol{\eta}_i,
\end{align}
\end{subequations} 
wherein, $\mu_{t,b}$ are the mobilities of the tracer and bath particles respectively, $U_\ext$ is an external potential acting on the tracer particle, $U$ is interacting potential between tracer and bath particles, and $V$ is potential between bath particles, $\bfxi_t$, $\bfxi_i$ and $\bfeta_i$ are white noises with zero means and unit variances, $\mathbf{f}_i$ is the self-propulsion force acting on bath particle $i$ with time correlation function $\langle \mathbf{f}_i(t) \mathbf{f}_j(t') \rangle = \frac{D_b}{\mu_b^2 \tau_b} \delta_{ij} e^{-|t-t'|/\tau_b} \mathbf{I}$, where $\mathbf{I}$ is the unit matrix 
, $\tau_b$ is the self-propulsion correlation time and $D_b$ serves as the amplitude of such force, with the same dimension as diffusivity.
We consider the situation that all potentials ($U$ and $V$) are square integrable, meaning that they have well defined and limited Fourier transforms: $U_k = \int U(\mathbf{r}) e^{i\mathbf{k}\cdot \mathbf{r}} \dif \mathbf{r}$, $V_k = \int V(\mathbf{r}) e^{i\mathbf{k}\cdot \mathbf{r}} \dif \mathbf{r} $.
\begin{figure}
  \centering
  \includegraphics[width=8.5cm]{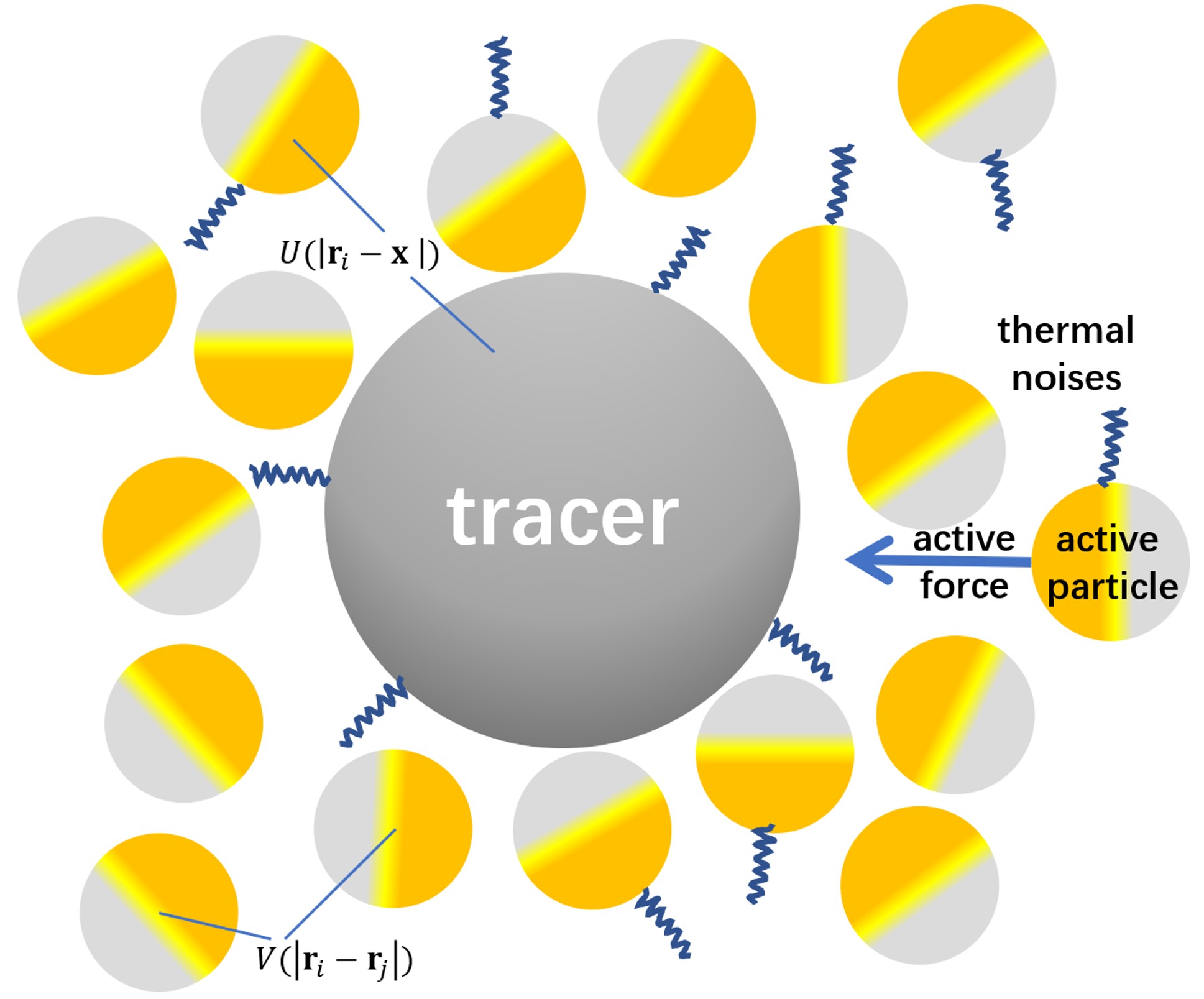}
    \caption{Schematic diagram of a tracer particle with coordinate $\bx$ in an active particle bath consisting of self-propulsion particles at $\br_i,\ i=1,2,...,N$. The tracer-bath interaction is $U(|\bx-\br_i|)$, and the bath-bath interaction is $V(|\br_i-\br_j|)$.}
\end{figure}

Inspired by the idea of Dean's works\cite{96JPA_Dean, 10PRL_Demery_DragForce}, we derive a self-consistent equation for the density profile of bath particles $\rho(\mathbf{r}, t)=\sum_{j=1}^N \delta (\mathbf{r}-\mathbf{r}_j(t))$,  (See details in Appendix \ref{sec:derive}), 
\begin{align}
    \frac{\partial}{\partial t} \rho(\mathbf{r}, t) =& \mu_b \nabla \cdot \left\{ \rho(\mathbf{r},t) \left[ \int \rho(\mathbf{r}',t) \nabla V(\vert \mathbf{r} - \mathbf{r}' \vert) \dif \mathbf{r}' \right.\right. \nonumber \\
    & + \nabla U(\vert \mathbf{r}-\mathbf{x} \vert) \Big]\Big\} +\mu_b T \nabla^2 \rho(\mathbf{r},t) \nonumber \\
    & + \nabla \cdot \left\{ \sqrt{\rho(\mathbf{r},t)} \left[ \boldsymbol{\xi}^T (\mathbf{r},t) + \boldsymbol{\xi}^A(\mathbf{r},t) \right] \right\} \label{eq:DeanR}
\end{align}
where the noise terms $\bfxi^{\{T,A\} }$ has correlation functions $\langle \bfxi^T(\br,t) \bfxi^T(\br ',t')\rangle = 2\mu_b T\delta(\br - \br ') \delta(t-t')\mathbf{I}$, $\langle \bfxi^A(\br,t) \bfxi^A(\br ',t')\rangle =\frac{D_b}{\tau_b} e^{-|t-t'|/\tau_b} \delta(\br - \br ')\mathbf{I}$.
In the Fourier space ($\delta \rho_k(t) = \int e^{i\br \cdot \bk} (\rho(\br,t) - \rho_0 )\dif \bk$), Eq.\eqref{eq:DeanR} has a formal solution,
\begin{multline}
    \delta\rho_k(t) = \int_{-\infty}^t e^{-(t-s)G_k} \Big\{ -\mu_b k^2 \rho_0 U_k e^{i\bk\cdot \bx_s} \\
    + \sqrt{\rho_0} i\bk \cdot \left[ \tilde{\bfxi}^T_k(s) + \tilde{\bfxi}^A_k(s) \right] \Big\} \dif s \label{eq:rhok}
\end{multline}
wherein $G_k = \mu_b k^2 (T+\rho_0 V_k)$ can be considered as a characteristic frequency (a typical illustration of this term is shown in Fig.\ref{fig:gap}(a)).%, and a mean-field approximation has been used $\sqrt{\rho(\br,t)} \approx \sqrt{\rho_0}$, $\rho_0 = N_b / V$. 
Noise terms $\tilde{\bfxi}^{\{T,A\} }_k(t)$ are the Fourier transform of $\bfxi^{\{T,A\} }$ respectively, with correlation functions
\begin{align}
    \left\langle \tilde{\bfxi}_k^T(t) \tilde{\bfxi}_{k'}^T(t') \right\rangle =& 2\mu_b T (2\pi)^3 \delta(\bk+\bk')\delta(t-t')\mathbf{I} \\
    \left\langle \tilde{\bfxi}_k^A(t) \tilde{\bfxi}_{k'}^A(t') \right\rangle =& \frac{D_b}{\tau_b} (2\pi)^3 \delta(\bk+\bk') e^{-|t-t'|/\tau_b} \mathbf{I}
\end{align}
wherein $\langle \cdots \rangle$ means the average over noises.

\subsection{Generalized Langevin equation}
To achieve an effective equation of motion of the tracer which does not contain any bath particle variables, one can use an identity $\nabla_x \sum_{i=1}^N U(\vert \br_i -\bx \vert) = - \frac{1}{(2\pi)^3} \int i\bk e^{-i\bk \cdot \bx} \rho_k U_k \dif^3 \bk$, then has a generalized Langevin equation for $\bx$, 
\begin{align}
    \dot{\bx}(t) =& -\mu_t \nabla_x U_{\ext}(\bx) + \mu_t \frac{1}{(2\pi)^3} \int i\bk e^{-i\bk \cdot \bx(t)} \delta\rho_k U_k \dif^3 \bk \nonumber \\
    & + \sqrt{2\mu_t T} \bfxi_t \nonumber \\
    =& -\mu_t \nabla_x U_\ext(\bx) - \int_{-\infty}^t \mathbf{K}(s,t) \dif s + \bfeta_T(t) + \bfeta_A(t) \nonumber \\
    & + \sqrt{2\mu_t T} \bfxi_t ,\label{eq:dotx}
\end{align}
where the memory kernel
\begin{align}
    \mathbf{K}(s,t) =& \frac{\mu_t \mu_b \rho_0}{(2\pi)^3} \int ik^2 \bk U_k^2 e^{-(t-s)G_k} e^{-i\bk \cdot (\bx(t)-\bx(s) ) } \dif^3 \bk \nonumber \\
    =& \frac{\mu_t \mu_b \rho_0}{(2\pi)^3} \int  \mathrm{Im} [e^{i\bk \cdot (\bx(t) - \bx(s))}] \bk k^2 U_k^2 e^{-(t-s)G_k}\dif^3 \bk
\end{align}
and the colored noise terms,
\begin{align}
    \bfeta_T(t) =& -\mu_t \frac{1}{(2\pi)^3} \int \bk e^{-i\bk \cdot \bx_t} \sqrt{ \rho_0} U_k \int_{-\infty}^t e^{-(t-s)G_k} \nonumber \\
    & \times \bk \cdot \tilde{\bfxi}_k^T (s) \dif s \dif^3 \bk \\
    \bfeta_A(t) =& -\mu_t \frac{1}{(2\pi)^3} \int \bk e^{-i\bk \cdot \bx_t} \sqrt{ \rho_0} U_k \int_{-\infty}^t e^{-(t-s)G_k}  \nonumber \\
    & \bk \cdot \tilde{\bfxi}_k^A (s) \dif s \dif^3 \bk
\end{align}
and their correlations are
\begin{align}
    \mathbf{G}_T(t,t') =& \left\langle \bfeta_T(t) \bfeta_T(t') \right\rangle \nonumber \\
    =& \frac{\mu_t^2 \rho_0 \mu_b T}{(2\pi)^3} \int  e^{-i\bk \cdot (\bx_t-\bx_{t'})} \mathbf{kk} k^2 U_k^2 \frac{ e^{-|t-t'|G_k}}{G_k} \dif^3 \bk \label{eq:CoEtaT_I} \\
    \mathbf{G}_A(t,t') =& \left\langle \bfeta_A(t) \bfeta_A(t') \right\rangle \nonumber \\
    =& \frac{\mu_t^2 \rho_0 D_b}{ (2\pi)^3} \int  e^{-i\bk \cdot (\bx_t-\bx_{t'})} \mathbf{kk} k^2 U_k^2 \nonumber \\
    &\times \frac{ \tau_b e^{-|t-t'|/\tau_b} - e^{-|t-t'|G_k}/  G_k }{\left[ (\tau_b G_k)^2 - 1 \right] } \dif^3 \bk \label{eq:CoEtaA_I}
\end{align}

\subsection{Path integral and effective diffusion}
To calculate the transport coefficients, one needs to calculate several correlation functions firstly. Considering the coupling between the tracer position and the colored noises $\bfeta_{A,T}$, we propose a path integral method to calculate them.
%The starting point is the Eq.\eqref{eq:dotx}. 

We consider a path of the tracer in the time interval $[t_i, t_f]$.
The \emph{partition function} of such trajectory can be written as 
\begin{multline}
    Z = \int \prod_t \delta\Big\{ \dot{\bx} + \mu_t \nabla U_\ext + \int_{-\infty}^t \mathbf{K}(s,t) \dif s - \bfeta_A - \bfeta_T \\
    - \sqrt{2\mu_t T} \bfxi_t \Big\} P[\bfxi_t] P[\bfeta_T] P[\bfeta_A] \calD \bx \calD \bfxi_t \calD \bfeta_A \calD \bfeta_T 
\end{multline}
Using the identity of delta function, $\delta(x) = \frac{1}{(2\pi)} \int e^{ipx} \dif p$, we also have
\begin{multline}
    Z=\int e^{ i\int \bp \cdot \left\{ \dot{\bx} + \mu_t \nabla U_\ext + \int_{-\infty}^t \mathbf{K}(s,t) \dif s - \bfeta_A - \bfeta_T - \sqrt{2\mu_t T} \bfxi_t \right\} \dif t } \\
    \times P[\bfxi_t] P[\bfeta_T] P[\bfeta_A] \calD \bx \calD \bp \calD \bfxi_t \calD \bfeta_A \calD \bfeta_T 
\end{multline}
where $\bp$ is an auxiliary real vector field.
Next, by utilizing $\bka{e^{au}} = e^{\frac{1}{2}a^2\bka{u^2}}$ for a Gaussian random variable $u$ with zero mean, we have the partition function as a function of the action,
\begin{equation}
    Z = \int e^{-S(\bx, \bp)} \calD \bx \calD \bp, \quad S(\bx, \bp) = S_0 (\bx, \bp) + S_{\rm int} (\bx, \bp)
\end{equation}
where 
\begin{equation}
    S_0 = -i\int \bp(t) \cdot \left[\dot{\bx}(t) + \mu_t \nabla_x U_{\ext} \right] \dif t + \mu_t T \int \lvert \bp(t) \rvert^2 \dif t
\end{equation}
for a free particle and
\begin{multline}
    S_{\rm int} = -i\int \bp(t) \cdot \left[\int_{-\infty}^t \mathbf{K}(s,t) \dif s \right] \dif t \\
    + \frac{1}{2} \iint \bp(t) \cdot [\mathbf{G}_T(t,s)+\mathbf{G}_A(t,s)] \cdot \bp(s) \dif t \dif s
\end{multline}
counts for the tracer-bath interaction.
After introducing the partition function over the trajectory, the average over any operator $A$ as function of $\bx(t),\ t\in[t_i,t_j]$ can be defined as 
\begin{equation}
    \langle A \rangle = \frac{ \int A e^{-(S_0+S_{\rm int})} \calD \bx \calD \bp }{Z} = \frac{ \langle A e^{-S_{\rm int}} \rangle_0 }{\langle e^{-S_{\rm int}} \rangle_0 }, \label{eq:NeqAve}
\end{equation}
where $\langle \cdots \rangle_0 = Z_0^{-1} \int (\cdots) e^{-S_0} \calD \bx \calD \bp$, corresponds to a tracer particle only affected by external potential, not any particle bath. %To further calculations, we need some correlators firstly. 
So far, the $e^{-S_{\rm int}}$ term is still too complicated to handle. A common treatment is the linear truncation when $S_{\rm int}$ is weak. Herein, we treat the tracer-bath interaction $U$ as a small perturbative quantity by assigning $U(r) = \sqrt{\epsilon} u(r)$, where $\epsilon$ is a dimensionless factor that scales the interacting strength, and can be used to the following perturbative expansion. 
The memory kernel $\mathbf{k}$ and correlators $G_{A,T}(t)$ are order 1 of $\epsilon$. Therefore, Eq.\eqref{eq:NeqAve} can be expand as 
%\begin{align}
%    \bka{A} =& \frac{\bka{A(1-S_{\rm int} )}_0 }{ \bka{1-S_{\rm int} }_0 }+ O(\epsilon^2) %\nonumber \\
%    =& \bka{A}_0 - \bka{AS_{\rm int}}_0 + \bka{A}_0 \bka{S_{\rm int}}_0 + \frac{1}{2} %\bka{AS_{\rm int}^2}_0 - \bka{AS_{\rm int}}_0 \bka{S_{\rm int}}_0 + \frac{1}{2} \bka{A}_0 %\bka{S_{\rm int}^2}_0 + O(\epsilon^3)
%\end{align}
\begin{equation}
    \bka{A} = \bka{A}_0 - \bka{AS_{\rm int}}_0 + \bka{A}_0 \bka{S_{\rm int}}_0 + O(\epsilon^2) \label{eq:Aneq}
\end{equation}
%To calculate $\bka{S_{\rm int}}_0$, one can use Wick's theorem. 
For studying the diffusion problem, we only need to consider the free particle situation, $U_\ext =0$. 
According to the symmetry, we have $\bka{\bp(t) e^{-i\bk \cdot [\bx(t)-\bx(s)]}}_0 = -i\bka{\bp(t) [\bx(t)-\bx(s)]}_0 \cdot \bk e^{-k^2 \mu_t T(t-s)} = 0$. 
Similarly, we also have 
\begin{multline}
    \bka{\bp(t) \cdot \mathbf{G}_{A,T}(t,s) \cdot \bp(t)}_0 \propto \Big\{ \bka{|\bp(t)|^2}_0 \\
    - \left( \bk \cdot \bka{ (\bx(t)-\bx(s))\bp(t) }_0 \right)^2 \Big\} e^{-k^2\mu_t T(t-s)} =0
\end{multline}
Therefore, 
\begin{equation}
    \bka{S_{\rm int} }_0 = 0
\end{equation}

The next step is to calculate the mean square displacement (MSD), i.e., to calculate $\bka{[\bx(t_f)-\bx(t_i)]^2}$ for a long time interval $t_f-t_i$. Then, the effective diffusion coefficient can be given as $D_\eff = \lim_{(t_f-t_i)\rightarrow{\infty}} \frac{1}{2d(t_f-t_i)}\bka{[\bx(t_f)-\bx(t_i)]^2} $, where $d$ is the dimension of the system. 
According to Eq.\eqref{eq:Aneq}, the key step of calculating the MSD is handling $\bka{[\bx(t_f)-\bx(t_i)]^2 S_{\rm int}}_0$, i.e. calculating the following two correlation functions, one is
\begin{align}
    & \bka{[\bx(t_f)-\bx(t_i)]^2 \bp(t) e^{-i\bk \cdot [\bx(t) - \bx(s)]} }_0 \nonumber \\
    =& -2i \bk \cdot \bka{[\bx(t)-\bx(s)][\bx(t_f)-\bx(t_i)]}_0 \nonumber\\
    & \bka{[\bx(t_f)-\bx(t_i)] \bp(t)}_0 e^{-k^2 \mu_t T (t-s)} \nonumber \\
    =& 4 \bk \mu_t T L([t_i, t_f]\cap[s,t]) \chi_{[t_i, t_f)}(t) e^{-k^2\mu_t T (t-s)} \nonumber \\
    =& 4 \bk \mu_t T [t-\max(t_i, s)] \chi_{[t_i, t_f)}(t) e^{-k^2\mu_t T (t-s)} \label{eq:xxp}
\end{align}
and the other is 
\begin{align}
    & \bka{  [\bx(t_f)-\bx(t_i)]^2 \bp(t) \bp(s) e^{-i\bk \cdot [\bx(t)-\bx(s)]} }_0 \nonumber \\
    =& \left\{ 4\mu_t T[t-\max(s,t_i)] \mathbf{kk} - 2\chi_{[t_i, t_f)}(s) \mathbf{1} \right\} \nonumber \\
    &\times e^{-k^2 \mu_t T(t-s)} \chi_{[t_i, t_f)}(t) \label{eq:xxpp}
\end{align}
for $t>s$ (the omitted mathematical details can be found in Appendix \ref{sec:math}). After the calculation in Eq.\eqref{eq:xxsapp}, we get 
\begin{multline}
    \bka{ [\bx(t_f)-\bx(t_i)]^2 S_{\rm int} }_0 
    = 2\mu_t T (t_f-t_i) \frac{\mu_t\mu_b\rho_0}{(2\pi)^3} \\
    \times \int \frac{k^4 U_k^2}{G_k(k^2\mu_t T + G_k)} \dif^3 \bk 
    + \frac{\mu_t^2 \rho_0 D_b}{(2\pi)^3} (t_f-t_i) \\
    \times \int \frac{2k^4 U_k^2}{(\tau_b G_k)^2-1} \left\{ \frac{\tau_b \mu_t T k^2 - 1}{(\mu_t T k^2 + \tau_b^{-1})^2} - \frac{\mu_t T k^2/G_k-1}{(\mu_t T k^2 + G_k)^2} \right\} \dif^3 \bk \\
    + o(t_f-t_i)
\end{multline}
where $G_k = \mu_b k^2 (T+\rho_0 V_k)$ reflects the bath properties including background temperature $T$, number density $\rho_0$ and interactions between bath particles. 
At last, taking the long-time limit of $t_f-t_i$, the $o(t_f-t_i)$ term can be neglected, the effective diffusion coefficient of a tracer in active bath is obtained
\begin{subequations}
\begin{equation}
D_\eff = \mu_t T \left\{ 1-\int_0^\infty g_T(k) \dif k + \int_0^\infty g_A(k) \dif k \right\}, \label{eq:deff}
\end{equation}
to the linear order, wherein
\begin{align}
    g_T(k) =& \frac{\mu_t\mu_b\rho_0}{6\pi^2} \frac{k^6 U_k^2}{G_k(k^2\mu_t T + G_k)} \\
    g_A(k) =& \frac{\mu_t\rho_0 D_b/T}{6\pi^2} \frac{ k^6 U_k^2}{(\tau_b G_k)^2-1} \left[ \frac{\mu_t T k^2/G_k-1}{(\mu_t T k^2 + G_k)^2} \right.\nonumber \\
    & -\left. \frac{\tau_b \mu_t T k^2 - 1}{(\mu_t T k^2 + \tau_b^{-1})^2} \right]
\end{align}
\end{subequations}
This is the main result of the present work.
In this formula, 1 in the brace denotes the bare diffusion of a free tracer particle. The second term in the brace denotes the ``passive part'' of the tracer-bath interaction which is always a negative contribution to effective diffusion and recovers the results in Ref.\cite{11PRE_Demery_TracerInFluids}. In the absent of activity, the FDT holds since the effective mobility of the tracer satisfies $\mu_\eff = D_\eff/T$ \cite{11PRE_Demery_TracerInFluids}. 
The third term is a pure ``active'' contribution on the diffusion, which is a positive contribution and explicitly gives the nontrivial dependence of $D_\eff$ on the bath parameters, $\rho_0$, $D_b$, $\tau_b$ and interactions $U(r)$ and $V(r)$.  For this linear truncation, $D_\eff$ is a linear function of $D_b$. However the dependence of $\tau_b$ and $\rho_0$ (which is also contained in $G_k$) are illegible, which require numerical calculations to determine (see Sec.\ref{sec:res}).
Further mathematical analysis of this expression is shown in Appendix \ref{sec:ana_deff}. 

\section{Simulation results}\label{sec:res}
In this section, we show numerical calculations of Eq.\eqref{eq:deff} with the persistent time of active force $\tau_b$ and the number density of bath particles $\rho$, at small activity $D_b$ region. 
Then, by comparing these results with computer simulations, the validity and applicability of the theory can be verified. 

In the present work, we choose $V(r)$ and $U(r)$ are both harmonic potentials, $V(r) = \frac{\kappa_b}{2}(r-\sigma_{bb})^2$ for $r<\sigma_{bb}$, and $U(r) = \frac{\kappa_t}{2}(r-\sigma_{tb})^2$ for $r<\sigma_{tb}$. We set $\sigma_{bb}$ is the unit of length, $\kappa_b\sigma_{bb}^2$ as the unit of energy, and $1 /(\mu_b \kappa_b)$ as the unit of time.  The common parameters are set as: $\sigma_{tb} = 2.0$,  $\kappa_t = 1.0$, $\mu_t = 0.333$, $T=1.0$. The other parameters are set as variables which are explicitly given in the following figures.
In computer simulations, we construct a three dimensional system with periodic boundary containing (1+4095) particles. The diffusion coefficient is calculated through a long time simulation ($\sim 10^8$ steps with $10^{-3}$ as the time step) and averaged over 20 samples with random initial configurations. 

\begin{figure}
    \centering
    \includegraphics[width=\columnwidth]{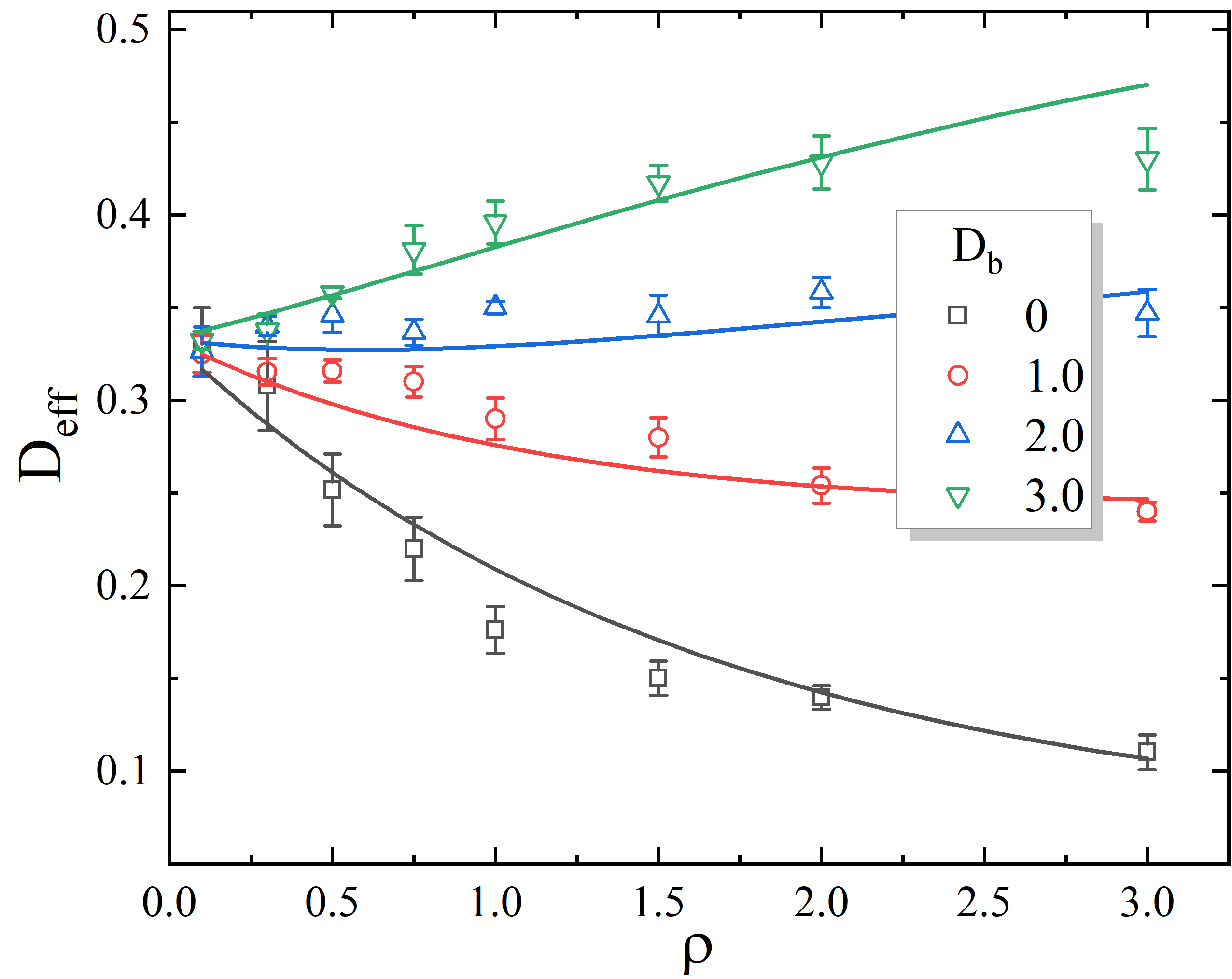}
    \caption{The dependence of the effective diffusion on the number density of active bath particle $\rho$. Dots and corresponding error bars are simulation results of $D_\eff$. Solid lines are the numerical calculations of $D_\eff$. Both results show that, with the increasing of activity $D_b$, effective diffusion $D_\eff$ gradually changes from monotonic decrease to monotonic increase with the number density of bath particle $\rho$, including a non-monotonic interval around $D_b=2.0$. Herein, $\tau_b$ is set as 0.1.}
    \label{fig:rho}
\end{figure}

Firstly, we focus on the contribution of bath number density $\rho$ on $D_\eff$, shown in Fig.\ref{fig:rho}. In Figs.(2) and (3), dots and corresponding error bars are the direct simulations, and lines are numerical calculation of Eq.\eqref{eq:deff}. In general, $D_\eff$ shows a diverse dependence on $\rho$. For small active force amplitude $D_b$, $D_\eff$ decreases with $\rho$ as the tracer's behavior in a passive particle bath. And for large activity situation, $D_\eff$ shows the opposite behavior. Consequently, one may expect that there is a moderate activity region that $D_\eff$ has a non-monotonic dependence on $\rho$.

\begin{figure}
    \centering
    \includegraphics[width=\columnwidth]{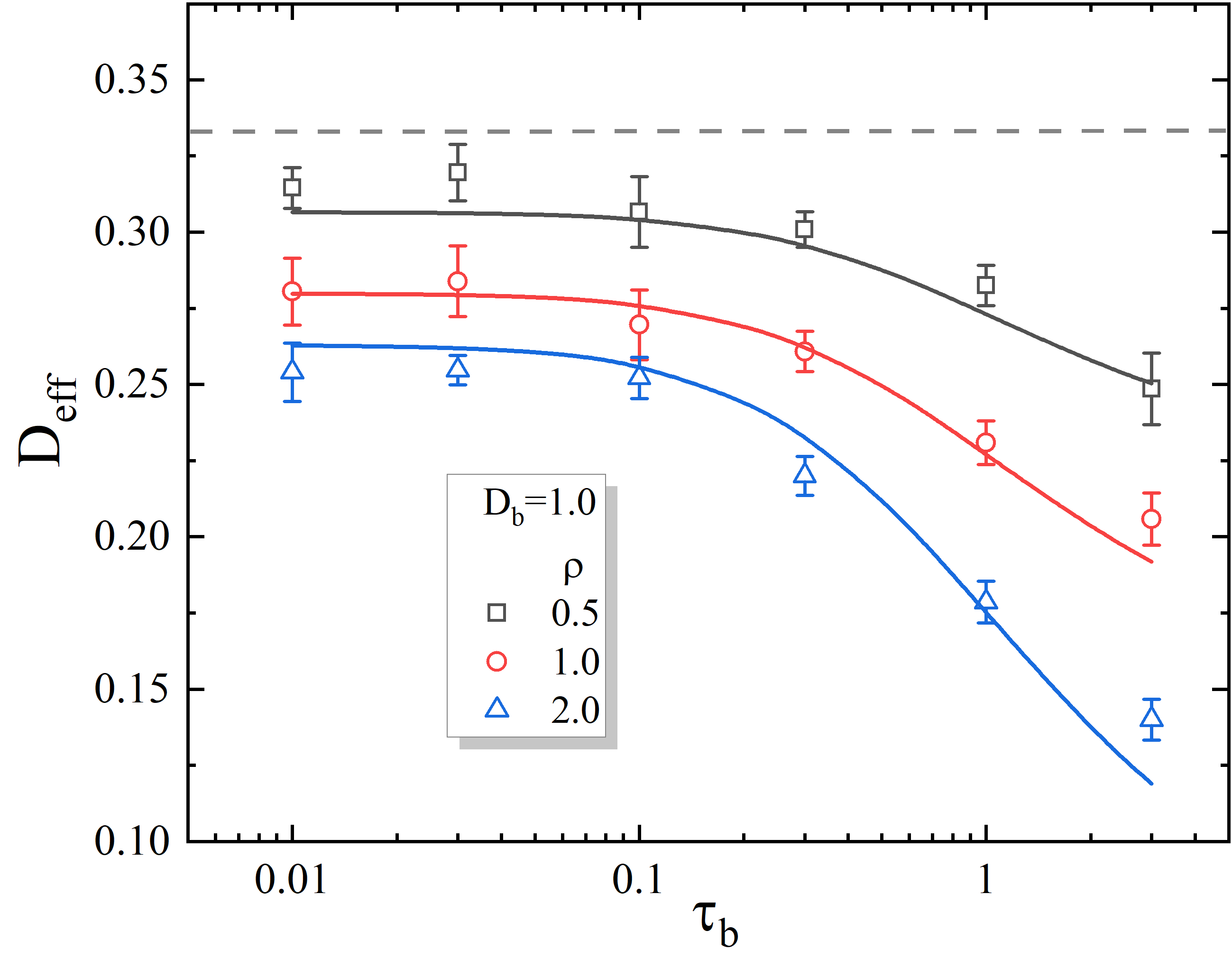}
    \caption{The dependence of the effective diffusion on the active force persistent time $\tau_b$. Dots and corresponding error bars are simulation results of $D_\eff$. Solid lines are the numerical calculations of $D_\eff$. Both show that $D_\eff$ monotonically decreases with the persistent time $\tau_b$. The horizontal dash line denotes the bare diffusion coefficient of the tracer particle.}
    \label{fig:taub}
\end{figure}

We then investigate the dependence of $D_\eff$ on persistent time $\tau_b$. According to Eq.\eqref{eq:deff}, $D_\eff$ decreases with $\tau_b$ since both $(G_k\tau_b)^2-1$ and $\frac{\tau_b\mu_t T k^2-1}{(\mu_t Tk^2+\tau_b^{-1})^2}$ increase with $\tau_b$ monotonically. This prediction has been confirmed in simulations, as shown in Fig.\ref{fig:taub}. Physically, when the persistent time of the active force tends to zero, the active OU particle can be reduced to an ordinary Brownian particle under a higher temperature. If the activity amplitude $D_b$ is constant, the longer $\tau_b$ means a larger deviation of equilibrium. Based on the results here, we might conclude that an active OU particle bath that more close to equilibrium, is more conducive to the tracer diffusion, when the activity amplitude is given.

\section{Conclusion}
This study aims to shed light on the intricacies of tracer diffusion in an active particle bath. By employing the generalized Dean's equation, incorporating the path integral method, and utilizing computer simulations, we characterize the impact of self-propulsion on the diffusion behavior of a passive tracer particle. 
In summary, the effective diffusion decreases with persistent time $\tau_b$, and exhibits a variety of dependencies on bath density, depending on $D_b$.
The obtained insights expand our understanding of collective dynamics and transport phenomena in non-equilibrium systems, with potential applications in diverse scientific disciplines.
For further studies, an extension of the active bath situation is straightforward since it has been confirmed to calculate the mobility of a tracer in particle bath \cite{11PRE_Demery_TracerInFluids,14NJP_GLE_Demery}. Additionally, after the effective mobility is achieved, the fluctuation-dissipation theorem can be further investigated to determine its validity or deviations with respect to the activity parameters.

\section*{Acknowledgement}
This work is supported by MOST(2022YFA1303100), NSFC(32090040,21833007).

\onecolumngrid
\appendix
\renewcommand{\thefigure}{S\arabic{figure}}
\setcounter{figure}{0}

\section{Derivation of Linearized Dean's equation\label{sec:derive}}
Although the general idea of the derivation has been given in Ref.\cite{23RPPFeng}, for the completeness of this article, we write it here as well.
Firstly, we briefly illustrate the derivation of Eq.\eqref{eq:dotx}. %The generalization to higher dimensional system is relatively straightforward. 
Consider a stochastic differential equation(SDE), $\frac{{\rm d}x}{{\rm d}t}=a(x,t)+b(x,t)\eta(t)$ where $\left\langle \eta(t)\right\rangle =0$ and $\left\langle \eta(t)\eta(t')\right\rangle =\delta(t-t')$ denotes the Gaussian white noise. 
Using the It\^o calculus, for any well behaved function $f(x)$, one has
\begin{equation}
    \frac{{\rm d}f(x)}{{\rm d}t}=\left[a(x,t)\partial_{x}f+\frac{b^{2}(x,t)}{2}\partial_{x}^{2}f\right]+\left(\partial_{x}f\right)b(x,t)\eta(t)
\end{equation}
Now substituting the Langevin equation \eqref{eq:LEs} of bath
particles for the SDE above, using the It\^o calculus,
we have an evolution equation for arbitrary function $g({\bf r}_{i})$
\begin{align}
\frac{{\rm d}g\left({\bf r}_{i}\right)}{{\rm d}t}= & \sqrt{2\mu_{b}T}\boldsymbol{\xi}_{i}\cdot\nabla_{i}g+\left(\nabla_{i}g\right)\left[-\mu_{b}\nabla_{i}\left(\sum_{j\neq i}V+U\right)+{\bf f}_{i}\right]+\mu_{b}T\nabla_{i}^{2}g\nonumber \\
= & \int{\rm d}{\bf r}\rho_{i}({\bf r},t)\left\{ \sqrt{2\mu_{b}T}\boldsymbol{\xi}_{i}\cdot\nabla g+\left(\nabla g\right)\Big[ \right. \nonumber \\
& \left. -\mu_{b}\nabla\left(\int{\rm d}{\bf r}'\rho({\bf r}',t)V\left(\left|{\bf r}-{\bf r}'\right|\right)+U\left(\left|{\bf r}-{\bf x}\right|\right)\right)+{\bf f}_{i}\Big]+\mu_{b}T\nabla^{2}g\right\} \nonumber \\
= & \int{\rm d}{\bf r}g({\bf r})\left\{ -\sqrt{2\mu_{b}T}\nabla\rho_{i}\cdot\boldsymbol{\xi}_{i}-\nabla\cdot\left[{\bf f}_{i}\rho_{i}({\bf r},t)\right]+\mu_{b}\nabla\rho_{i}({\bf r},t)\right.\nonumber \\
 & \cdot\left.\left[\nabla\left(\int{\rm d}{\bf r}'\rho({\bf r}',t)V\left(\left|{\bf r}-{\bf r}'\right|\right) + U\left(\left|{\bf r}-{\bf x}\right|\right)\right)\right]+\mu_{b}T\nabla^{2}\rho_{i}({\bf r},t)\right\} 
\end{align}
wherein the bath density $\rho\left({\bf r},t\right)=\sum_{i}\rho_{i}\left({\bf r},t\right)=\sum_{i}\delta\left({\bf r}-{\bf r}_{i}\left(t\right)\right)$ is applied. On the other hand, we can also write $\frac{{\rm d}g({\bf r}_{i})}{{\rm d}t}=\int\frac{\partial\rho_{i}({\bf r},t)}{\partial t}g({\bf r}){\rm d}{\bf r}$.
Considering the arbitrariness of function $g({\bf r})$, we have 
\begin{multline}
    \frac{\partial\rho_{i}\left({\bf r},t\right)}{\partial t}=-\sqrt{2\mu_{b}T}\nabla\rho_{i}\cdot\boldsymbol{\xi}_{i}-\nabla\cdot\left[{\bf f}_{i}\rho_{i}\right]+\mu_{b}\nabla\rho_{i} \\
    \cdot\left[\nabla\left(\int{\rm d}{\bf r}'\rho({\bf r}',t)V\left(\left|{\bf r}-{\bf r}'\right|\right)+U\left(\left|{\bf r}-{\bf x}\right|\right)\right)\right]+\mu_{b}T\nabla^{2}\rho_{i}
\end{multline}
for single particle density and the collective density profile
\begin{multline}
\frac{\partial\rho\left({\bf r},t\right)}{\partial t}=-\sum_{i}\sqrt{2\mu_{b}T}\nabla\cdot\left[\boldsymbol{\xi}_{i}\rho_{i}\right]-\sum_{i}\nabla\cdot\left[{\bf f}_{i}\rho_{i}\right] \\
+\mu_{b}\nabla\rho\cdot\left[\nabla\left(\int{\rm d}{\bf r}'\rho({\bf r}',t)V\left(\left|{\bf r}-{\bf r}'\right|\right)+U\left(\left|{\bf r}-{\bf x}\right|\right)\right)\right]+\mu_{b}T\nabla^{2}\rho\label{eq:den_not_cns}
\end{multline}
notice that this equation is not closed for the moment, due to the first two noise terms. To achieve Eq.\eqref{eq:dotx}, we need to introduce the noise field to replace these terms, meanwhile keeping the correlation properties invariant. 
Let $\chi_{1}(t)=-\sum_{i}\sqrt{2\mu_{b}T}\nabla\cdot\left[\boldsymbol{\xi}_{i}(t)\rho_{i}({\bf r},t)\right]$,
$\chi_{2}(t)=-\sum_{i}\nabla\cdot\left[{\bf f}_{i}(t)\rho_{i}({\bf r},t)\right]$,
the time correlation function of these terms are
\begin{align*}
\left\langle \chi_{1}(t)\chi_{1}(t')\right\rangle = & 2\mu_{b}T\delta(t-t')\sum_{i=1}^{N}\nabla_{r}\cdot\nabla_{r'}\left[\rho_{i}({\bf r},t)\rho_{i}({\bf r}',t')\right]\\
= & 2\mu_{b}T\delta(t-t')\nabla_{r}\cdot\nabla_{r'}\left[\sum_{i=1}^{N}\rho_{i}({\bf r},t)\delta({\bf r}-{\bf r}')\right]\\
= & 2\mu_{b}T\delta(t-t')\nabla_{r}\cdot\nabla_{r'}\left[\rho\left({\bf r},t\right)\delta\left({\bf r}-{\bf r}'\right)\right]
\end{align*}
\begin{align*}
\left\langle \chi_{2}(t)\chi_{2}(t')\right\rangle = & \frac{D_{b}}{\tau_{b}}e^{-\left|t-t'\right|/\tau_{b}}\sum_{i=1}^{N}\nabla_{r}\cdot\nabla_{r'}\left[\rho_{i}({\bf r},t)\rho_{i}({\bf r}',t')\right]\\
= & \frac{D_{b}}{\tau_{b}}e^{-\left|t-t'\right|/\tau_{b}}\nabla_{r}\cdot\nabla_{r'}\left[\sum_{i=1}^{N}\rho_{i}({\bf r},t)\delta({\bf r}-{\bf r}')\right]\\
= & \frac{D_{b}}{\tau_{b}}e^{-\left|t-t'\right|/\tau_{b}}\nabla_{r}\cdot\nabla_{r'}\left[\rho\left({\bf r},t\right)\delta\left({\bf r}-{\bf r}'\right)\right]
\end{align*}
Now we introduce two global noise fields, $\chi_{1}'\left({\bf r}, t\right) = \nabla\cdot\left[\sqrt{\rho\left({\bf r},t\right)} \boldsymbol{\xi}^{T}\right]$ and $\chi_{2}'\left({\bf r},t\right) = \nabla\cdot\left[\sqrt{\rho\left({\bf r},t\right)}\boldsymbol{\eta}_{f}\right]$,
where the correlations are $\left\langle \boldsymbol{\xi}^{T}\left({\bf r},t\right)\boldsymbol{\xi}^{T}\left({\bf r}',t'\right)\right\rangle =2\mu_{b}T\delta\left(t-t'\right)\delta\left({\bf r}-{\bf r}'\right){\bf 1}$
and $\left\langle \boldsymbol{\xi}^{A}\left({\bf r},t\right) \boldsymbol{\xi}^{A}\left({\bf r}',t'\right)\right\rangle = \frac{D_{b}}{\tau_{b}}\delta\left({\bf r}-{\bf r}'\right){\bf 1}$ respectively. 
It is worth noting that the noises $\chi_{1}$ and $\chi_{1}'$ have the same time correlation function, also true for $\chi_{2}$ and $\chi_{2}'$. Now we find the replacement and Eq.\eqref{eq:den_not_cns} becomes 
\begin{multline}
\frac{\partial\rho\left({\bf r},t\right)}{\partial t}=\nabla\cdot\left[\sqrt{\rho\left({\bf r},t\right)}\boldsymbol{\xi}^{T}\right]+\nabla\cdot\left[\sqrt{\rho\left({\bf r},t\right)}\boldsymbol{\xi}^{A}\right] \\
+\mu_{b}\nabla\rho\cdot\left[\nabla\left(\int{\rm d}{\bf r}'\rho({\bf r}',t)V\left(\left|{\bf r}-{\bf r}'\right|\right)+U\left(\left|{\bf r}-{\bf x}\right|\right)\right)\right]+\mu_{b}T\nabla^{2}\rho 
\end{multline}
i.e. Eq.\eqref{eq:DeanR}, which is a self-consistent equation. 

Although we use this replacement technique (to the second order moment of the noise fields) to simply the time evolution equation of collective density, this equation is still to hard to solve. 
For a system that the density fluctuation is small, we can further use a mean-field assumption that $\sqrt{\rho\left({\bf r},t\right)}\approx\sqrt{\rho_0}$
where $\rho_0=N/V$ is the averaged number density. This approximation
is more precise for higher densities and used to simplify the interaction terms, $\rho({\bf r})\int{\rm d}{\bf r}'\rho({\bf r}')\nabla V\left(\left|{\bf r}-{\bf r}'\right|\right) \approx \rho_0\rho({\bf r})*\nabla_{r}V({\bf r})$ and $\rho({\bf r})\nabla U\left(\left|{\bf r}-{\bf x}\right|\right) \approx \rho_0 \nabla U \left(\left|{\bf r}-{\bf x}\right|\right)$,
which is $i{\bf k}e^{i{\bf k}\cdot{\bf x}}\bar{\rho}U_{k}$ in Fourier space. Using the Fourier transform $\delta\rho_{k}(t)=\int e^{i{\bf k}\cdot{\bf r}}\left[\rho\left({\bf r},t\right)-\rho_0 \right]{\rm d}{\bf r}$, we have 
\begin{equation}
    \frac{\partial}{\partial t} \delta\rho_{k}\left(t\right) = -\mu_{b}k^{2}\left[\left(T+ \rho_0 V_{k}\right)\delta\rho_{k}\left(t\right) + \rho_0 U_{k}e^{i{\bf k}\cdot{\bf x}}\right]+\sqrt{\rho_0}i{\bf k} \cdot \left[\tilde{\boldsymbol{\xi}}_{k}^{T} \left(t\right)+\tilde{\boldsymbol{\xi}}_{k}^{A}\left(t\right)\right]
\end{equation}
where $V_{k}$ , $U_{k}$, $\tilde{\boldsymbol{\xi}}_{k}^{T}$ and
$\tilde{\boldsymbol{\xi}}_{k}^{A}$ are Fourier transforms of $V(r)$,
$U(r)$, $\boldsymbol{\xi}^{T}$ and $\boldsymbol{\xi}^{A}$ respectively.
The formal solution of this function is the Eq.\eqref{eq:rhok} in the main text.

\section{Calculate correlations with path integral}\label{sec:math}
For convenience, we set $\bx(t=0)=0$, and consider the situations for $t \ge 0$ and $U_\ext=0$.
According to the symmetry, we immediately we have $\bka{\bx(t)-\bx(t') }_0 = 0$, $\bka{\bp(t)}_0 = 0$. 
Using the identity $\int \frac{\delta}{\delta \bx(s)} [\bp(t)e^{-S_0}] \calD \bx \calD \bp \equiv 0$, we have $\bka{i\bp(t) \dot{\bp}(s)} = 0$, i.e. $\bka{\bp(t)\bp(s)}_0$ is a constant. Considering the action does not correlate with $\bp$ at different time, we must have $\bka{\bp(t)\bp(s)}_0 = 0$. 
Also, using identity  $\int \frac{\delta}{\delta \bp(s)} [\bp(t)e^{-S_0}] \calD \bx \calD \bp \equiv 0$, we have $\bka{i\bp(t) \dot{\bx}(s) }_0 = -\delta(t-s)\mathbf{I} + 2\mu_t T\bka{\bp(t) \bp(s)}_0$. In the case of free particle, we have
\begin{equation}
\bka{[\bx(t)-\bx(t')] \bp(s)}_0 = i\mathbf{I} \chi_{[t',t)}(s),    
\end{equation}
where $\chi_{[t',t)}(s) =  \Bigg\{ \begin{array}{cc}
    1 & s\in [t',t) \\
    0 & {\rm otherwise}
\end{array} $, and using $\int \frac{\delta}{\delta \bp(s)}[\bx(t)e^{-S_0}] \calD \bx \calD \bp \equiv 0$, we have 
\begin{equation}
\bka{ [\bx(t)-\bx(t')][\bx(s)-\bx(s')] }_0 = 2\mu_t T \mathbf{I} L([t',t] \cap [s',s]).    
\end{equation}
where the $L$ function is the length of the time interval. 

After Eq.\eqref{eq:xxp} and \eqref{eq:xxpp}, we obtain
\begin{align}
    & \bka{ [\bx(t_f)-\bx(t_i)]^2 S_{\rm int} }_0 \nonumber \\
    =& \bka{ -i[\bx(t_f)-\bx(t_i)]^2 \int \bp(t) \cdot \left[ \int^t \mathbf{K}(s,t) \dif s \right] \dif t }_0 \nonumber \\
    & + \bka{\iint_{t>s} [\bx(t_f)-\bx(t_i)]^2 \bp(t) \cdot [\mathbf{G}_T(t,s)+\mathbf{G}_A(t,s)] \cdot \bp(s) \dif s \dif t }_0 \nonumber \\
    =& \frac{\mu_t \mu_b \rho_0}{(2\pi)^3} \int_{t_i}^{t_f} \dif t \int^{t} \dif s 4\mu_t T [t-\max(t_i,s)] \int k^4 U_k^2 e^{-(t-s)G_k} e^{-k^2\mu_t T (t-s)} \dif^3 \bk + \frac{\mu_t^2 \rho_0 \mu_b T}{(2\pi)^3} \nonumber \\
    & \times \int_{t_i}^{t_f} \dif t \int^t \dif s \int \dif^3 \bk \left\{ 4\mu_t T[t-\max(s,t_i)] k^2 - 2\chi_{[t_i, t_f)}(s)  \right\} e^{-k^2 \mu_t T(t-s)} k^4 U_k^2 \frac{e^{-(t-s)G_k}}{G_k} \nonumber \\
    & + \frac{\mu_t^2 \rho_0 D_b}{(2\pi)^3} \int_{t_i}^{t_f} \dif t \int^t \dif s \int \dif^3 \bk \left\{ 4\mu_t T[t-\max(s,t_i)] k^2 - 2\chi_{[t_i, t_f)}(s)  \right\} \nonumber \\
    & \times e^{-k^2 \mu_t T(t-s)} k^4 U_k^2 \frac{\tau_b e^{-(t-s)/\tau_b} - e^{-(t-s)G_k}/G_k}{(\tau_b G_k)^2 - 1} \nonumber \\
    =& \mu_t T \frac{\mu_t\mu_b\rho_0}{(2\pi)^3} \int \frac{ 4 k^4 U_k^2}{(k^2\mu_t T+G_k)^2} \dif^3 \bk  (t_f-t_i)  + \mu_t T \frac{\mu_t\mu_b\rho_0}{(2\pi)^3} \int  \frac{2 k^4 U_k^2 (\mu_t T k^2 - G_k)}{G_k(k^2\mu_t T + G_k)^2} \dif^3 \bk (t_f-t_i) \nonumber \\
    +& \frac{\mu_t^2 \rho_0 D_b}{(2\pi)^3} \int \frac{2k^4 U_k^2}{(\tau_b G_k)^2-1} \left\{ \frac{\tau_b \mu_t T k^2 - 1}{(\mu_t T k^2 + \tau_b^{-1})^2} - \frac{\mu_t T k^2/G_k-1}{(\mu_t T k^2 + G_k)^2} \right\} \dif^3 \bk (t_f-t_i)  + o(t_f-t_i) \label{eq:xxsapp}
\end{align}
In this equation, the first term after the last equal sign is contributed by the memory kernel, the second and the third terms are due to the bath induced noises, wherein the second one labels the contribution of thermal noises of bath particle, and the last one is pure active noise contribution. 
Interestingly, the first two terms can be reduced to
\begin{equation*}
    \mu_t T  (t_f-t_i) \frac{\mu_t\mu_b\rho_0}{(2\pi)^3} \int  \frac{2 k^4 U_k^2 (\mu_t T k^2 /G_k + 1)}{ (k^2\mu_t T + G_k)^2} \dif^3 \bk = 2\mu_t T  (t_f-t_i) \frac{\mu_t\mu_b\rho_0}{(2\pi)^3} \int \frac{k^4 U_k^2}{G_k(k^2\mu_t T + G_k)} \dif^3 \bk,
\end{equation*}
which is positive definite, meaning that the passive particle bath is always a negative contribution (due to $-\bka{AS_{\rm int}}_0$ term) to the tracer diffusion.
%As for the active part, it does not have the property of positive or negative definite. And a case-by-case analysis is required.

\section{Analysis of $D_\eff$} \label{sec:ana_deff}
\begin{comment}
\begin{multline}
    D_{\eff} = \mu_t T \Bigg\{ 1-\frac{\mu_t\mu_b\rho_0}{3(2\pi)^3} \int \frac{k^4 U_k^2}{G_k(k^2\mu_t T + G_k)} \dif^3 \bk \\
    + \frac{\mu_t \rho_0 D_b/T}{3(2\pi)^3} \int \frac{ k^4 U_k^2}{(\tau_b G_k)^2-1} \left[  \frac{\mu_t T k^2/G_k-1}{(\mu_t T k^2 + G_k)^2} - \frac{\tau_b \mu_t T k^2 - 1}{(\mu_t T k^2 + \tau_b^{-1})^2} \right] \dif^3 \bk \Bigg\} \label{eq:deff}
\end{multline}
The reason why we choose the sign in Eq.\eqref{eq:deff} is to ensure that the integrals over $g_{T,A}$ are positive.
\end{comment}
\begin{figure}
    \centering
    \includegraphics[width=15cm]{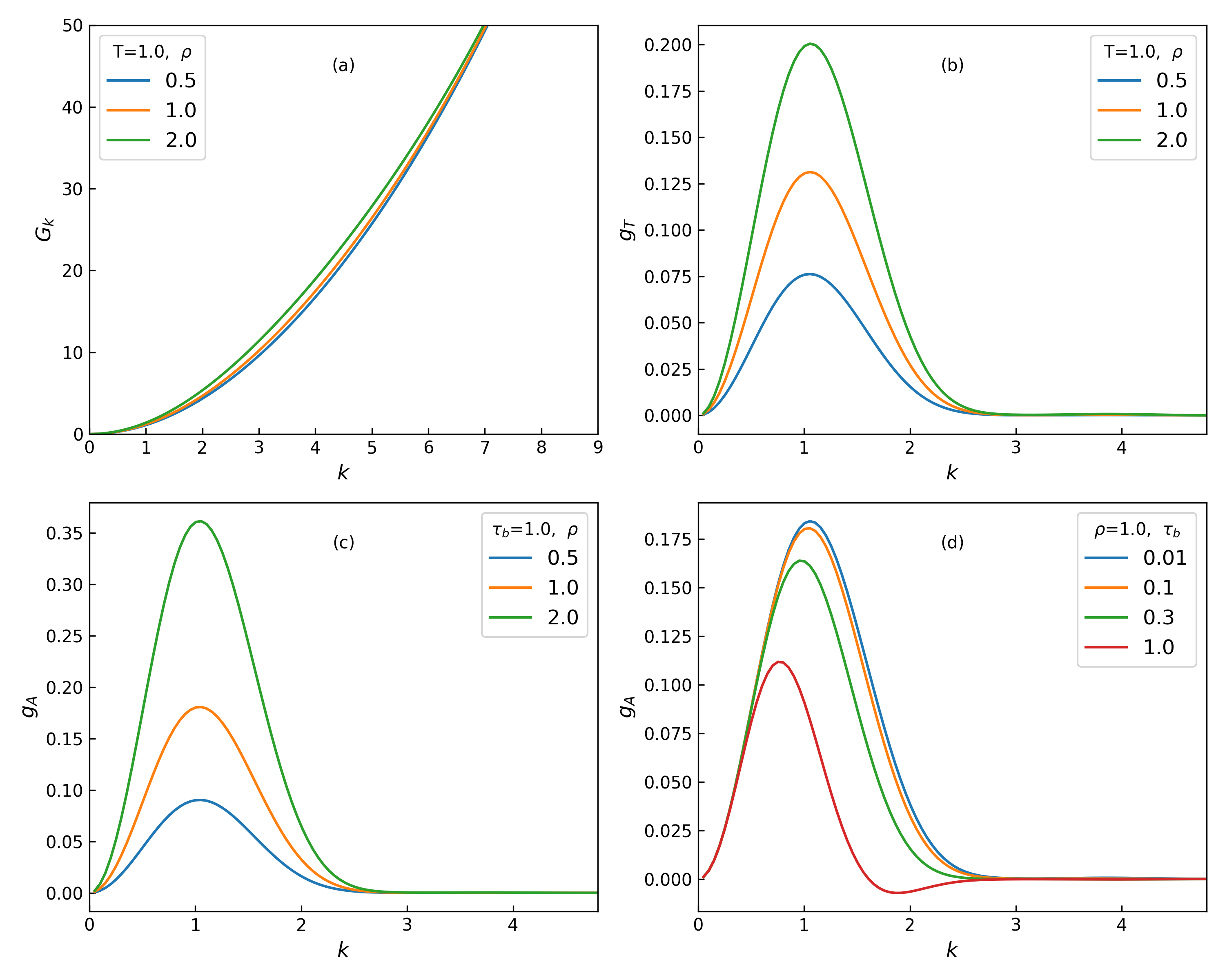}
    \caption{Visualization of the integrated functions in Eq.\eqref{eq:deff}. (a) is the plot for the characteristic frequency $G_k$; (b) is the passive part contribution of $D_\eff$ with three different number density $\rho$; (c) and (d) are the active part contribution; (c) is for constant persistent time $\tau_b$ and different $\rho$, and (d) shows the effects of $\tau_b$. }
    \label{fig:gap}
\end{figure}
From the perspective of practical application, we hope that the expression \eqref{eq:deff} can be simplified, such as, by avoiding the computation of this integral.
To give an intuitive image of two functions, we draw Fig.\ref{fig:gap} to show these function under several different control parameters. Herein, we firstly show the dependence of the characteristic frequency of the bath $G_k$ on bath number density in Fig.\ref{fig:gap}(a). As we expected, $G_k$ grows with $k$ and approaches $\mu_b T k^2$ asymptotically. In Fig.\ref{fig:gap}(b) and (c), we plot the functions $g_T(k)$ and $g_A(k)$ under three typical $\rho$s. And further examines the influence of $\tau_b$ on $g_A(k)$ as shown in Fig.\ref{fig:gap}(d). 
In noticing these curves, most of them are extremely similar to the Gaussian bell curve, except for the large $\tau_b$ cases in (d). 
Mathematically, the reason is that the numerators of $g_{A,T}$ contain $U_k$ term which highly closes to the Gaussian function, and eventually lead to $g_{A,T} \propto k^2 U_k^2$ for large $k$s. For instance, in 3D system, for the harmonic potential $U(r)=\frac{\kappa_t}{2}(|r|-R_t)^2$ we used in simulations in the present work, we have $U_k = \frac{4\pi \kappa_t }{k^3} \left[ (\frac{6}{k^{2}}-2R_t^{2}) \sin (kR_t) - \frac{6R_t}{k}\cos (kR_t) \right] \approx \frac{8\pi \kappa_t R_t^5}{15} [1+\frac{5(kR_t)^4}{3528}+\dots] \exp[-\frac{(kR_t)^2}{14}] $, % + \frac{215(kR_t)^6}{1629936} \approx \frac{8\pi \kappa_t R_t^5}{15} [1-\frac{(kR_t)^2}{14} + \frac{(kR_t)^4}{252}  -\frac{(kR_t)^6}{33264}+\dots]
and for Gaussian potential $U(r)=\kappa_t \exp [-(\frac{r}{R_t})^2]$, $U_k = \kappa_t \pi^{\frac{3}{2}} R_t^3 \exp [-\frac{(kR_t)^2}{4}]$ is an exact Gaussian function.
This fact suggests that one might be able to use the saddle point approximation to replace the integration, $\int g(k) dk \approx g(k^*) \sqrt{-\frac{2\pi g(k^*)}{g''(k^*)}}$ where $k^*$ satisfies $g'(k^*)=0$.
Although this Gaussian approximation analysis does not ultimately provide a concise and universally applicable expression for $D_\eff$, it still provides a feasible method to facilitate the calculation of $D_\eff$.
%In addition, it has been noticed that in Fig.\ref{fig:gap}(d), the integral over $g_A(k)$ may leads to a negative result, which may be unphysical. In simulations, we have never observed that the additional activity leads to a smaller effective diffusion coefficient. In our opinion, this mathematical solution is derived from the previous linear truncation Eq.\eqref{eq:Aneq}. For the present system, $D_\eff$ is still reliable for not very large $\tau_b$ such as $\tau_b \leq 1.0$. 

\end{document}